\documentclass[11pt]{article}
\usepackage{moriond,epsfig,epsf}
\def\G1915{GRS 1915+105}
\def\X1550{XTE J1550--564}
\def\J1655{GRO J1655--40}

\def\etal{et al. }
\def\ie{{\em i.e. } }

\def\ergcms{erg/cm$^{2}$/s }

\begin{document}
\vspace*{4cm}
\title{RXTE Observations of XTE J1550--564 during its 2000 outburst}

\author{J. Rodriguez$^1$,  S. Corbel$^2$, E. Kalemci$^3$ \& J.A. Tomsick$^3$}
\address{$^1$DSM/DAPNIA/Service d'Astrophysique, CEA Saclay, 91191 Gif-sur-Yvette, France\\
$^2$Universit\'e Paris VII  and Service d'Astrophysique, CEA, CE-Saclay. 91191
Gif sur Yvette, France.\\
$^3$Center for Astrophysics and Space Science, University of California San Diego, USA}

\maketitle\abstracts{
We report on timing and spectral observations of
the microquasar XTE J$1550-564$ during its 2000 outburst made by  the 
Rossi X-ray Timing Experiment (RXTE). We study the spectral properties of 
the source during both the 
rise to the outburst and its decline. We observe transitions from
a low hard state to an intermediate state, and vice versa, 
before the source returns to quiescence. We show that the first transition
 likely reflects a change in the relative importance of the emitting media,
 instead of a change in the total accretion rate, contrary to the second 
transition.
 We investigate  the temporal properties of the source, and follow the 
evolution of a low frequency  Quasi Periodic Oscillations (QPO).}

\section{Introduction}
Soft X-ray transients (SXT) are accretion powered binary systems, 
hosting a compact object (either a neutron star or a black hole), which 
spend most of their life in quiescence, and are detected in the X-rays as 
they undergo episodes of outburst. 
Their X-ray spectra can usually be divided into two parts, representing 
different physical processes acting in the close vicinity of the accreting 
object. The soft X-rays are usually  taken as the thermal signature from an
 optically thick geometrically thin accretion disk, whereas the hardest 
emissions are understood as the inverse Compton scattering of the soft photons 
on relativistic electrons present in a optically thin coronal medium. 
Depending on whether the electrons have a thermal velocity distribution or not,
 this ``hard tail'' can be characterized by the presence or absence of an 
exponential cut off, at a certain threshold energy. 
It is thus possible to characterize their spectra with the relative shapes and 
strength of the different components,
 giving birth to a classification in terms of 5 canonical spectral states 
 (see Goldwurm, these proceedings).
 Some of these systems have been observed to produce radio jets giving 
birth to the so-called ``microquasar'' class of 
object (see Mirabel, these proceedings).\\
\indent \X1550 was first detected by the {\em{All Sky Monitor}} (ASM) on board RXTE, on
 September 7, 1998 (Smith 1998). Soon after, it exhibited the brightest flare 
($\sim 7$ Crab) observed with RXTE. The whole outburst  ended  $\sim 9$ 
months later in June 1999. 
Homan \etal (2001), based on a timing analysis, showed that the X-ray 
state transitions  over the whole outburst needed an additional parameter 
to the accretion rate $\dot{M}$, in order to be explained. 
The mass function  leads to $M_{Compact}=10.56 \pm 1.5$ $ M_\odot$ for the 
compact object (Orosz \etal  2002), implying the presence of a black hole. 
The latest estimations from optical observations give a prefered distance 
of $\sim 5.3-5.9$ kpc (Orosz \etal  2002). The inclination
 $i$ to the line of sight is $70.8^{\circ} \leq i \leq 75.4^{\circ}$. \\
\indent Low and high frequency QPOs have been 
detected in some PCA observations (resp. $\sim 0.01-20$, and $185-285$ Hz), 
making \X1550 the fourth black hole source where high frequency QPO have been 
reported (Sobczak \etal  2000; Remillard \etal  1999).
Radio observations performed during the 2000 outburst have shown the 
 presence of a compact jet during the low state (LS), which is quenched 
 during the intermediate state/very high state (IS/VHS)(Corbel \etal  2001).

\section{Outburst overview}
\begin{figure}[htbp]
\epsfig{file=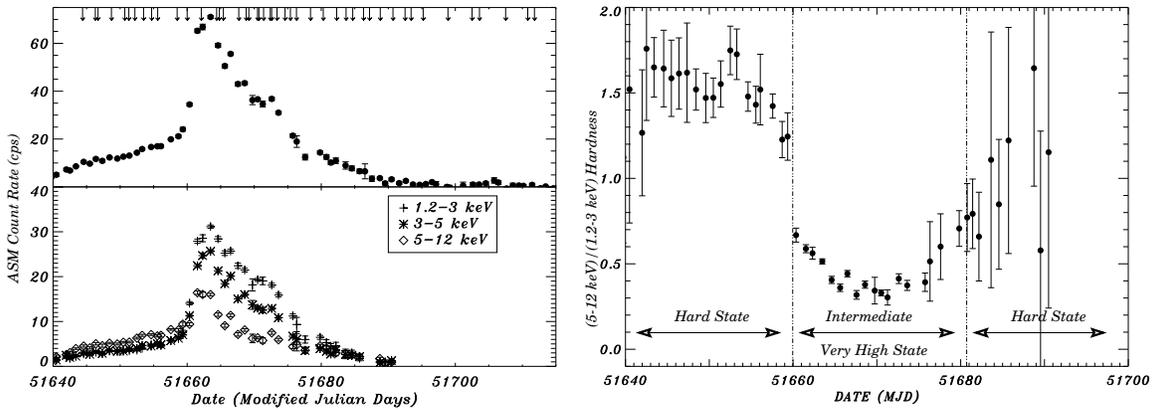,width=\columnwidth}
\caption{{\it Left} ASM light curves of the outburst Y axis is the count rate. 
Upper pannel 1.2-12 keV energy range; lower pannel 1.2-3,3-5,5-12 keV energy 
ranges. {\it Right} $\frac{5-12}{1.2-3}$ keV hardness ratio. Vertical lines 
indicate the dates of state transitions.}
\label{fig:asmlite}
\end{figure}

\indent On 2000 April 6, \X1550 became active (Smith \etal  2000), undergoing a 
new episode of outburst, which ended $\sim 70$ days after. 
The RXTE/ASM ($1.2-12$ keV) light curve is shown on Fig. 
\ref{fig:asmlite} together with the light curves in the three bands \ie
 $1.2-3$ keV, $3-5$ keV, $5-12$ keV.

The shape of the light curve increases slowly during an initial stage before, 
it becomes fast rise exponential decay-like (usually called FRED envelope) 
during the remaining period. From the $5-12$/$1.2-3$ keV
Hardness ratio (Fig. \ref{fig:asmlite}), we can identify two distinct spectral
 states (which we identify from the spectral studies of PCA+HEXTE spectra 
presented in section \ref{sec:spec}) as a LS, and a IS/VHS. 

\section{Spectral Results}
\label{sec:spec}
\begin{figure}[htbp]
\epsfig{file=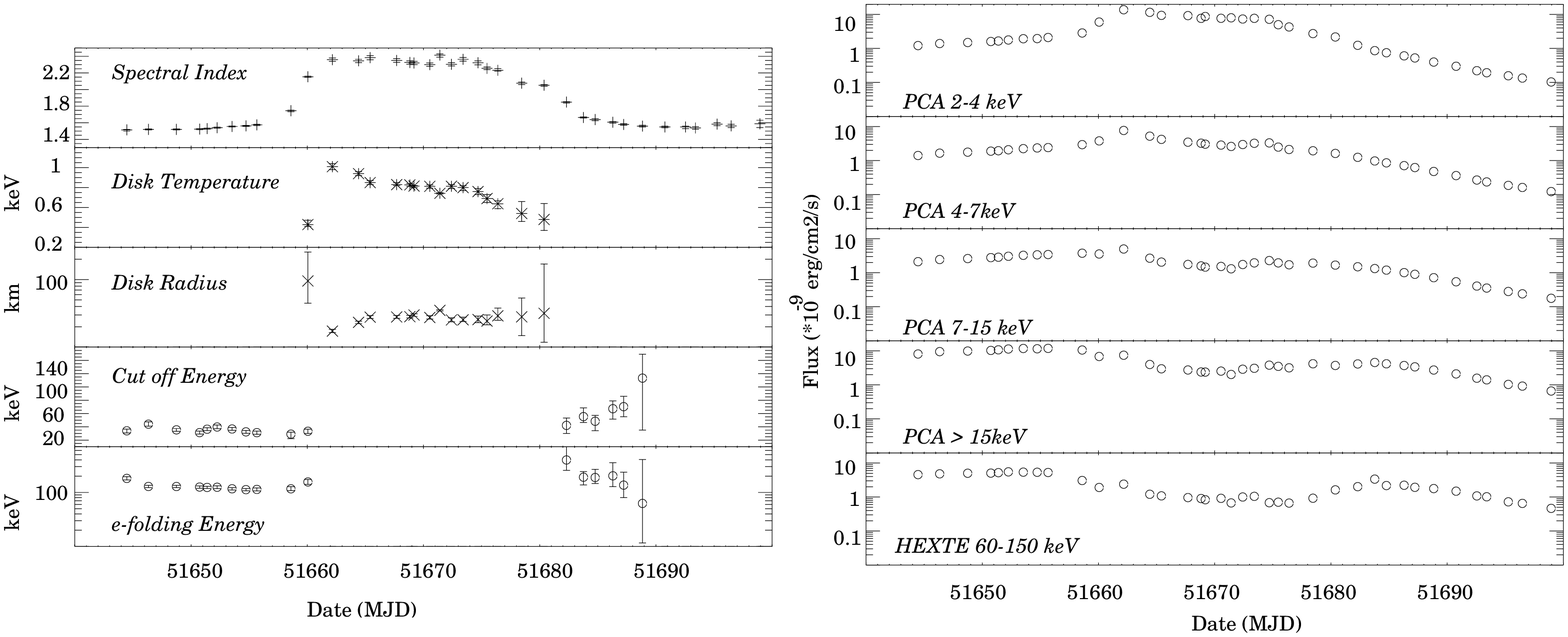,width=\columnwidth}
\caption{{\bf{Left: }} light curves of the spectral parameters returned from the
 fits. From top to bottom, powerlaw spectral index, disk Temperature (keV), 
disk inner radius (km assuming d=6 kpc, and $0^{\circ}$ inclination), cut-off
 energy (keV), and e-folding energy (keV). Vertical lines show the dates of 
state transitions.
{\bf{Right: }} multicolor light curves. Vertical axis is in terms 
of $10^{-9}$ \ergcms.}
\label{fig:spectral}
\end{figure}

\indent We fitted the PCA+HEXTE
 spectra between $3$ and $150$ keV with a model consisting of interstellar 
 absorption, a smeared Iron edge at $\sim 7$ keV, and a power law. 
An exponential cut-off is needed during the initial LS. During the IS/VHS,
 this cut-off disappears, and  a soft excess becomes visible in the spectra. 
This feature is modeled with a multi-color disk-blackbody. The 
 light curves of the returned parameters are shown in Fig. \ref{fig:spectral}.

The absence of a (detectable) thermal component, and the presence 
of a strong hard X-ray tail with an  exponential cut-off, favors an 
interpretation where most of the X-rays are produced from the comptonization 
of soft ($\leq 0.1$ keV) photons on thermal electrons in the corona.
 In the standard comptonization picture, the e-folding energy
  is thought to be close to the electron temperature.
 The slight decrease of the folding energy (from $181.8_{-11.2}^{+15.8}$ keV, 
on MJD 51644 to $116.6_{-9.9}^{+11.6}$ keV on MJD51658, Fig. \ref{fig:spectral}
 left) then indicates that the cooling of the corona becomes more efficient. 
This could be understood if the accretion
 disk is approaching the black hole, implying that, as the inner radius 
decreases, both its inner temperature and its flux increase, 
as suggested by the future state transition. 
The transition itself is rather abrupt, and manifests at the same time 
by the disappearance of the cut-off, a steepening of the power law shape, and
 and sudden increase of the disk color temperature up to $\sim 1$ keV, before
 declining to $\sim 0.8$ keV. Its constancy other $\sim 9$ days may
 indicate that the disk reaches a physical limit. The inner
 radius is found between $25-30$ km over 13 observations (assuming 6kpc, and
 $0^{\circ}$ inclination). Its relative constancy suggests that it reaches 
its last stable orbit. This last point is further confirmed by the
 detection of high frequency QPOs ($251-276$ Hz, Miller \etal  2001) almost 
the highest values as yet observed in this source ($285$ Hz, Remillard \etal 
 1999). This would suggest  that the black hole in \X1550 is 
rotating, leading to a  spin value $\sim0.47$ ($M_{BH}=8.5 M_\odot$) 
$\leq a \leq \sim 0.76$ ($M_{BH}=11.5 M_\odot$), 
in good agreement with the upper value of 0.6 found from the analysis of 
high frequency QPO  (Remillard \etal 2002b).
\begin{figure}[htbp]
\centering
\epsfig{file=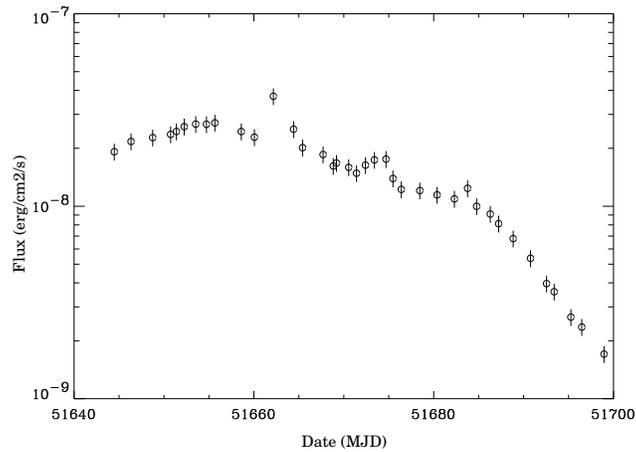,width=9cm}
\caption{2-150 keV light curve of the outburst}
\label{fig:tot}
\end{figure}
 After MJD 51680  the source slowly returns into a LS. 
Although the transition to the final low state is not so sharp as in 
the initial  stage, there is a clear evolution, in terms of spectral 
parameters, between MJD 51680 and MJD 51682, where the spectrum gets harder 
and manifests a cut-off at high energies (Tomsick, Corbel \& Kaaret, 2001). 
From MJD 51682 to 51698 the source is in a LS, which first shows 
the presence of an exponential cut off at high energy (51682-51688), 
while the following dates do not show any  cut-off up to
 150 keV. After MJD 51698, the observations become contaminated by both
 the Galactic ridge emission and the close outbursting transient pulsar
 XTE J1543-568.\\
\indent The total 2-150 keV luminosity of the source is plotted in fig. 
\ref{fig:tot}, showing in particular the relative constancy of the flux during
 the first state transition (around MJD 51660).
As discussed in Zhang \etal (1997) this almost constant luminosity more likely
 reflects a change in the relative importance of the emitting media 
(as suggested by the  independent evolution of the spectral parameters,
 and the energy dependent behaviors of the light curves in fig. 
\ref{fig:spectral} right), rather than a pure  change in the total accretion 
rate.

\section{Evolution of a Low Frequency QPO}
\begin{figure}[htbp]
\epsfig{file=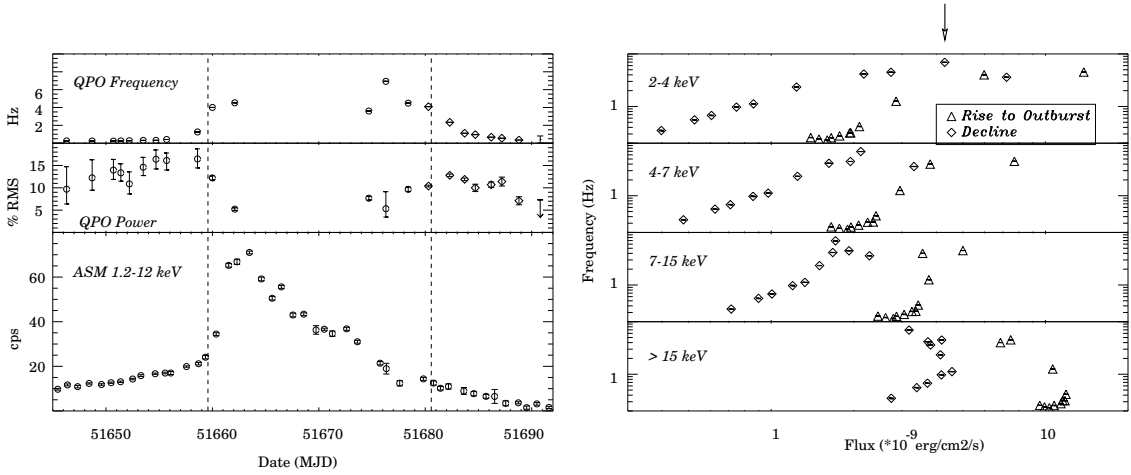,width=\columnwidth}
\caption{{\bf{Left: }}evolution of the low frequency QPO frequency (top), 
power (middle) vs.
 time. The ASM light curve is shown in the lower panel. {\bf{Right: }} QPO 
frequency vs. flux in four different PCA energy ranges. The arrow indicates 
the place of the turnover of the correlation discussed in the text.}
\label{fig:QPO}
\end{figure}

The power spectra of the source over the whole outburst show the presence
 of a low frequency QPO with a frequency that seems to correlate with the flux
 (Fig. \ref{fig:QPO}, Left). 
Indeed, from Fig. \ref{fig:QPO} we can see that the frequency of the QPO 
increases while the ASM flux increases, and vice-versa. In order to further
 investigate this behavior we plotted the evolution of the frequency vs. the
 flux in 4 PCA energy bands on Fig. \ref{fig:QPO} (Right). We can see that
 the QPO frequency correlates better with the flux in the lower energy
 ranges. The frequency presents a linear increase with the flux before it
 reaches a plateau and then inverts (the transition is marked with an arrow
 on fig. \ref{fig:QPO}). If the evolution of the soft flux follows 
the behavior of 
the disk radius as suggested by our spectral analysis, then the LFQPO 
appears to be somehow related to the evolution of the disk. This would then be 
consistent with the theoretical predictions of the Accretion Ejection 
Instability (AEI) suggested by Tagger \& Pellat (1999). 
In that case, part of the accretion energy and angular momentum of 
the inner disk are extracted and transported toward the corotation radius by 
a spiral wave. The QPO would then represent the spiral as it rotates 
in the disk, at $10\%-30\%$ of the Keplerian frequency at the inner radius. 
 The QPO frequency is thus expected to scale as $R_{in}^{-\frac{3}{2}}$.  
The plateau and the inversion of the slope (Fig. \ref{fig:QPO} right) 
 occur at high soft flux ($\geq 2\times10^{-9}$ \ergcms) at times where the 
disk reaches the highest temperatures, and contributes the most to the energy 
spectra, meanwhile the power law contribution has decreased significantly. 
If the disk is close to its last stable orbit as suggested from our spectral 
analysis, then the decrease of the QPO frequency after the plateau 
(Fig. \ref{fig:QPO} right) has the same origin as the frequency-radius 
inversion  of the correlation  observed in the case of \J1655 
(Sobczak \etal 2000; Rodriguez \etal 2002b), which has been shown, in the 
framework of the AEI, to be a signature of general relativistic modifications 
of the spiral rotation curve (Varni\`ere, Rodriguez \& Tagger, 2002).

\section{Conclusion}
The source has transited from an initial LS into a IS/VHS, then back into a
 LS before slowly returning to quiescence. The whole outburst evolution can be
 understood in terms of the evolution of an accretion disk which gets 
 brighter as it approaches the compact object.\\
We have studied the frequency behavior of a low frequency QPO over the outburst
, and shown that the QPO might be the signature of a 
Keplerian rotating spiral in the disk, whose rotation curve follows the 
evolution of the accretion disk, and 
becomes modified by general relativistic effects when the disk is close to 
its last stable orbit as expected from the theoretical predictions of the 
AEI.\\

%\indent From the correlation with the multiwavelength behavior of the source
% we have shown that the suggestion presented in Corbel \etal (2001), that
% the second IR-optical flare may be the synchrotron tail of the compact
% jet they observed in the radio domain. This is consistent with the X-ray 
%excess we detect around MJD 51690, which, given the simultaneity with the IR-
%optical flare, may be associated with the jet. The disappearance of the 
% coronal cut-off coincident with the excess, may further confirm the different
% origin of this spectral component.  

\end{document}